\documentclass[conference]{IEEEtran} 
\usepackage{amsmath}
\usepackage{latexsym} 
\usepackage{epsfig}

\newtheorem{proposition}{Proposition}
\newtheorem{definition}{Definition} 

\newtheorem{theorem}{Theorem} 
\newtheorem{lemma}{Lemma}

\def\Proof{\par\noindent{\bf Proof:}\indent}
\def\EOP{\hfill$\Box$\par\vskip1em}

\def\PROG#1{$\mathcal{#1}$}

\def\TT#1{\texttt{#1}}

\begin{document}

\title{Void Traversal for Guaranteed Delivery in Geometric Routing}
\date{}

\author{Mikhail Nesterenko\thanks{This research was
supported in part by DARPA contract OSU-RF\#F33615-01-C-1901
and by NSF CAREER Award 0347485}
and Adnan Vora \\
Computer Science Department \\
Kent State University \\
Kent, OH, 44242 \\
mikhail@cs.kent.edu, avora@cs.kent.edu \\
}

\date{}

\maketitle

\begin{abstract}
Geometric routing algorithms like GFG (GPSR) are lightweight, scalable
algorithms that can be used to route in resource-constrained ad hoc
wireless networks.  However, such algorithms run on planar graphs
only.  To efficiently construct a planar graph, they require a
unit-disk graph. To make the topology unit-disk, the maximum link
length in the network has to be selected conservatively. In practical
setting this leads to the designs where the node density is rather
high. Moreover, the network diameter of a planar subgraph is greater
than the original graph, which leads to longer routes. To remedy this
problem, we propose a void traversal algorithm that works on arbitrary
geometric graphs. We describe how to use this algorithm for geometric
routing with guaranteed delivery and compare its performance with GFG.
\end{abstract}

\section{Introduction} \label{SecIntro}
Geometric routing is a promising approach for message transmission in
ad hoc wireless networks.  Unlike traditional ad hoc routing,
geometric routing algorithms have no control messages or routing
tables, the nodes maintain no information about data messages between
transmissions and each data message is of constant size. Hence,
geometric routing is quite scalable. Geometric routing is particularly
appropriate for wireless sensor networks. Networked sensors usually
have limited resources for routing information, yet many applications
for sensor networks require ad hoc configurations of large scale.

In geometric routing, each node knows it own geographic coordinates.
Each node is also aware of the coordinates of other nodes and links in
a part of the network around it. The coordinates are either obtained
from GPS receivers or a localization algorithm
\cite{Bulusu:2004:SCL}. The source node of a data message has the
coordinates of the target but the source does not know the complete
route to it. The source selects one of its neighbors and sends the
message to it.  After getting the message the receiver forwards the
message further.  The objective of the algorithm is to deliver the
message to the target.

The simplest version of a geometric routing algorithm is \emph{greedy}
routing. In greedy routing the source, as well as each intermediate
node, examines its neighboring nodes and forwards the message to the
one that is the closest to the target. Unfortunately, the greedy
routing algorithm fails if the message recipient is \emph{local
minimum}: all its neighbors are further away from the target than the
node itself. In \emph{compass} routing \cite{CCCG99*51}, the next hop
is selected such that the angle between the direction to the next hop
node and to the destination is minimal. However, compass routing
livelocks even on planar graphs.

GFG \cite{Bose01} (also known as GPSR \cite{MOBICOM00*243}) is the
first algorithm that guarantees delivery of the message. The algorithm
is designed for planar graphs. It uses greedy routing. To get out of a
local minimum GFG sequentially traverses the faces of the planar graph
that intersect the line between the source and target. GFG assumes
that the original communications graph is a unit-disk graph. For this
graph, the authors construct its Gabriel subgraph. This subgraph
preserves the connectivity of the original graph and it can be
constructed locally by each node. Datta et al \cite{DSW02} propose
further improvements to GFG.  Kuhn et al \cite{DialM`02*24} present a
similar algorithm with asymptotically optimal worst-case performance.

\ \\ \textbf{Our contribution}. \ \ The guaranteed delivery geometric
routing algorithms, of which we are aware, have the following
shortcoming.  Their local minimum avoidance part runs over a planar
graph. The efficient construction of a planar graph requires that the
original graph is unit-disk. As recent work \cite{G+02b} demonstrates
this assumption is not realistic for connection topologies formed by
common networked sensors such as Berkeley motes
\cite{Hill:2002:MWP}. Motes' radio propagation patters prove to be
quite intricate.

In this paper we demonstrate how to carry out geometric routing with
guaranteed delivery on arbitrary non-planar graphs. The foundation of
our approach is a void traversal algorithm. We describe algorithms
VOID-1 and VOID-2 that incorporate void traversal for routing across
the whole network. We also present GVG --- an algorithm that uses
greedy routing and void traversal to get out of local minima.  The
void traversal is based on each node storing the network topology of
its neighborhood. The neighborhood size needs to meet a condition we
call \emph{intersection semi-closure}.  Most practical geometric
graphs meet this condition such that the storage requirements for each
node are independent of the size of the network and rather small.

We compared the performance of VOID-2 with FACE-2 which is the
foundation of GFG. We used randomly generated graphs as a base for
comparison. A large number of graphs did not have unit-disk subgraphs.
Hence FACE-2 cannot be run. On graphs where FACE-2 and VOID-2 are run,
depending on the graph generation parameters, the paths selected by
VOID-2 were 35-75\% shorter. The memory storage requirements for
VOID-2 were independent of the network size and only modestly
increased compared to those of FACE-2.

\ \\ \textbf{Paper organization.} \ \ This paper is organized as
follows. We give definitions and describe our notation in Section
\ref{SecPrelim}. We present our void traversal algorithm in Section
\ref{SecTraversal}. In Section \ref{SecRouting} we describe how this
algorithm can be used for geometric routing. In Section \ref{SecEval}
we describe the results of the performance comparison of the void
traversal versus face traversal algorithms.  Section
\ref{SecConclusion} concludes the paper.

\section{Preliminaries} 
\label{SecPrelim}

\noindent
\textbf{Notation.}\ \ We model a sensor network as a geometric graph.
A geometric graph $G = (V,E)$ is a set of \emph{nodes}
(\emph{vertices}) $V$ on a Euclidean plane connected by \emph{edges}
$E$. The number of nodes is $n=|V|$. Such graph is \emph{planar} if
its edges intersect only at vertices. \emph{Void} is a region on a
plane such that any two points of this region can be connected by a
curve that does not intersect any of the edges of the graph. A
boundary of a void contains the segments of the edges adjacent to the
void. In every finite graph there is one unbounded \emph{external}
void.  A void in a planar graph is \emph{face}. Observe that the
boundary of a face forms a simple cycle. Moreover, each edge of a
planar graph borders at most two faces. An edge of a non-planar graph
may contain the segments of the borders of arbitrarily many faces.

\emph{Neighborhood} $N(u)$ of a node $u$ is a subgraph of $G$. A
\emph{neighborhood relation} \PROG{N}$(G)$ over graph $G$ associates
each vertex of $V$ with a subgraph of $G$.  Denote $(u,v) \in E$ an
edge between nodes $u$ and $v$. Denote $path(u,v)$ a path from $u$ to
$v$. Denote also $|u,v|$ the Euclidean distance between $u$ and $v$.

%
% mention somewhere that the neighborhood means that the node knows
% the geometric coordinates of the ``neighbors''
%

\ \\ \noindent
\textbf{Intersection semi-closure.}

\begin{definition} \label{DefSemiclosure}
A neighborhood relation \PROG{N}$(G)$ over a geometric graph $G$ is
\emph{$d$-incident edge intersection semi-closed} (or just
\emph{$d$-intersection semi-closed}) if, for every two intersecting
edges $(u,v)$ and $(w,x)$, either:
\begin{itemize}
\item there exist $path(u,w)$, $path(u,x)$, either one no more than
$d$ hops and $(w,x)$, $path(u,w)$ and $path(u,x)$ belong to $N(u)$; or
\item there exist $path(v,w)$, $path(v,x)$, either one no more than
$d$ hops and $(w,x)$, $path(v,w)$, and $path(v,x)$ belong to $N(v)$.
\end{itemize}

\end{definition}

The attractive feature of the intersection semi-closed neighborhood
relation is that for any edge $(u,v)$, the information about every
edge that intersects $(u,v)$ as well as how to reach such edge is
contained in the union of $N(u)$ and $N(v)$. Observe that for a
particular graph, depending on the value of $d$, such a relation
may not exist. However, for any graph there always exists
a $n$-incident edge semi-closed relation.

To illustrate Definition \ref{DefSemiclosure}, we apply it to
unit-disk graphs.  \emph{Unit-disk graph} is a geometric graph where
two nodes $u$ and $v$ are connected by an edge if an only if $|u,v|
\leq 1$.

\begin{figure} 
\center
\epsfig{figure=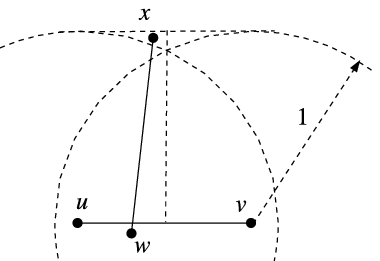,width=6cm,clip=} 
\caption{Edge $(w,x)$ belongs to $N(u)$ in a unit-disk graph}
\label{FigUdisk} 
\end{figure}

\begin{lemma}\label{UDiskNeighborhood}
A neighborhood relation over a unit-disk graph is 2-intersection
semi-closed if for every node $u$ and every edge $(w,x)$ such that
$|u,w| \leq 1$ and $|u,x| \leq 2/\sqrt{3}$ it follows that $(u,w) \in
N(u)$.
\end{lemma}

\Proof By definition \PROG{N}$(G)$ is 2-intersection semi-closed if
for every edge $(u,v)$ and every edge $(w,x)$ intersecting $(u,v)$,
and the route to $w$ and $x$ are either in $N(u)$ or $N(v)$. These
routes are no longer than 2 hops.

Let $N(u)$ and $N(v)$ be according to the conditions of the lemma.
Since $G$ is unit-disk, if $|u,w| \leq 1$ and $|u,x| \leq 1$ then
$(u,w) \in E$ and $(u,x) \in E$. According to the conditions of the
theorem, both of these edges as well as $(w,x)$ itself belong to
$N(u)$. If this is the case, then both the route to the endpoints and
the edge itself belong to $N(u)$. Moreover, these routes are at most
1-hop long. Hence, the definition of intersection semi-closure is
satisfied. Similar argument applies to $v$.

Suppose that $|u,x| > 1$ and $|v,x| > 1$. Refer to Figure
\ref{FigUdisk} for illustration. Since $(w,x)$ intersects $(u,v)$, $x$
could not be further away from $(u,v)$ than $1$. Hence, $|u,w| < 1$
and $|v,w| < 1$. Since $G$ is a unit-disk, $(u,w) \in G$ and $(v,w)
\in G$. By the conditions of the lemma $(u,w) \in N(u)$ and $(v,w) \in
N(v)$. Suppose that $x$ is closer to $u$ than to $v$. Then, the the
distance between $x$ and $u$ is at most $2/\sqrt{3}$.  By the
conditions of the lemma $(w,x) \in N(u)$. Which means that $N(u)$
contains the edge $(w,x)$ and, since $N(u)$ also contains $(u,w)$,
$N(u)$ contains the routes to both $w$ and $x$. Moreover, the
$path(u,x)$ is just 2-hops long: $(u,w),(w,x)$. The argument for $v$
is similar. The lemma follows.  \EOP

\ \\
\noindent
\textbf{Geometric routing.}\ \ 
%
% This is nice but too formal
%
%
% Given a connected geometric graph
% $G=(V,E)$, a neighborhood relation \PROG{N}$(G)$ and a pair of nodes
% $\{s,t\} \in V$ a \emph{geometric routing algorithm with guaranteed
% delivery} is a function $ALG(m_i,N(u_i),s,t) \rightarrow
% (m_{i+1},u_{i+1})$ such that if $u_1=s$, then there is a finite $k$
% such that $u_k=t$.  The first element in the range of $ALG$ ---
% $m_{i+1}$ is a \emph{message}, the second --- $u_{i+1}$ is \emph{next
% hop}.
% 
% We are interested in routing algorithms with the following
% restrictions on its arguments: (a) there is a constant $c$ such that
% for every node $u$ in the neighborhood relation \PROG{N}$(G)$ another
% node $v$ belongs to $N(u)$ only if $|u,v| \leq c $; (b) the size
% of any message is in $O(\log n)$ bits.
% 
% The attractive features of such algorithm is that it uses constant
% size messages only and the nodes store neither global nor
% destination-specific routing information. Moreover, in a geometric
% graph with constant maximum density, the neighborhood size of each
% node is also bounded. 
%
%
Consider a connected geometric graph $G=(V,E)$, a neighborhood
relation \PROG{N}$(G)$ over it and a pair of nodes $(s,t) \in V$.
Source $s$ has a message to transmit to target $t$. The source knows
the coordinates of the target. The message may be transmitted via
intermediate nodes. Each node may potentially add data to the
transmitted message. In the sequel we ignore the payload of the
message and assume that it contains only the routing information.

A routing algorithm specifies a procedure for intermediate node
selection.  A \emph{geometric routing algorithm with guaranteed
delivery} ensures the eventual message delivery under the following
two constraints: each node $u$ receiving the message, selects the next
node only on the basis of $N(u)$ and the contents of the message $u$
received; the message size is independent of network size.  Observe
that the nodes are not allowed to keep any information about the
transmitted message after it has been forwarded.

\section{Void Traversal Algorithm}
\label{SecTraversal}

\begin{figure} 
\center \epsfig{figure=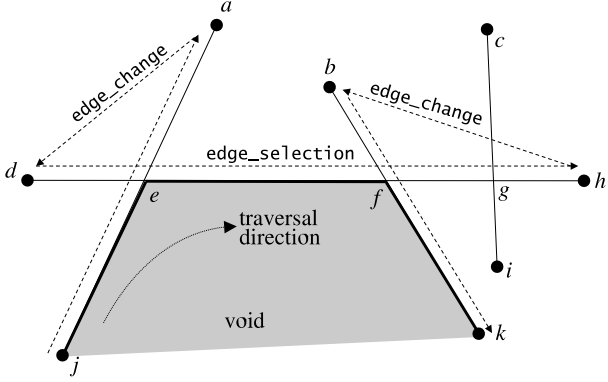,width=8cm,clip=}
\caption{Traversal of void $jefk$.}
\label{FigTraversal} 
\end{figure}

\noindent
\textbf{Overview.}\ \ The algorithm traverses an internal void
clockwise following the segments of the edges that comprise the border
of the void. The external void is traversed counter-clockwise.  Refer
to Figure \ref{FigTraversal} for the illustration. Given an edge, for
example $(d,h)$, that contains the segment of the void's border, the
algorithm has to select the next edge. For this the algorithm needs to
determine the ends of the segment of $(d,h)$ that borders the
void. One of the ends --- $e$ is the intersection of the previous edge
$(a,j)$ and $(d,h)$. The other end --- $f$ is the point of the
intersection of $(d,h)$ and another edge $(b,k)$ such that $f$ is
closest to $e$ in the direction of the traversal of the void.

Recall that in the intersection semi-closed neighborhood relation, the
union of the neighborhoods of $d$ and $h$ contains every edge
intersecting $(d,h)$ as well as a route to this edge. Hence, to
accomplish its objective the algorithm has to examine the
neighborhoods of $d$ and $h$.

\ \\ \noindent
\textbf{Details.}\ \
The algorithm uses two types of messages: \TT{edge\_change}, and
\TT{edge\_selection}.  \TT{edge\_change} contains: previous edge,
current edge and the direction of the traversal of the current
edge. \TT{edge\_selection} message contains: previous edge, and
suggested next edge.

When a node $d$ receives \TT{edge\_change} message from node $a$, it
determines the intersection point $e$ between the previous edge
$(a,j)$ and the current edge $(d,h)$. Notice that the intersection
point may be $d$ itself.  Then $d$ examines $N(d)$ to find the edge
whose intersection point is the closest to $e$ in the direction of
traversal. This edge is the suggested next edge.  Note that the graph
is intersection semi-closed and $N(d)$ may not contain some of some
edges that intersect $(d,h)$. For example: $(b,k) \not \in N(d)$.

Hence, $d$ selects $(c,i)$ as the suggested next edge. Edge $(c,i)$
intersects $(d,h)$ at point $g$.  If $d$ does not find any edge that
intersects $(d,h)$, it keeps the suggested next edge field empty.
Node $d$ sends \TT{edge\_selection} to the other node $h$ incident to
the current edge.

When node $h$ receives \TT{edge\_selection} from $d$, $h$ compares the
suggested next edge $(c,i)$ that $h$ receives from $d$ with the edges
in $N(h)$.  If $h$ finds an edge $(b,k) \in N(h)$ intersecting $(d,h)$
whose intersection point $f$ is closer to $e$ than $g$, then $h$ makes
$(b,k)$ the suggested next edge. Otherwise, the suggested next edge
remains the same. If neither $d$ nor $h$ find the suggested next edge
in this manner, $h$ considers the edges incident to itself. Node $h$
selects the edge nearest to $(d,h)$ counter-clockwise.

Node $h$ forms an \TT{edge\_change} message. This message contains
$(d,h)$ as the previous edge and $(b,k)$ as the current edge.  From
the information contained in \TT{edge\_selection} that $h$ receives
from $d$, $h$ is able to determine the direction of the traversal of
$(d,h)$. The traversal direction is included in
\TT{edge\_change}. 

After composing \TT{edge\_change}, $h$ sends this message to either
$b$ or $k$.  Since the graph is intersection semi-closed, $N(h)$ may
not contain the route to either nodes. In this case $h$ returns the
message to $d$ and $d$ forwards it to the appropriate node.

The above discussion is summarized in the following theorem.

\begin{theorem}\label{TraversalOK}
The Void Traversal Algorithm correctly traverses an arbitrary void in
a geometric graph with intersection semi-closed neighborhood relation.
\end{theorem}

\section{Using Void Traversal \\ to Guarantee Delivery}
\label{SecRouting}

\textbf{VOID-1 and VOID-2.}\ \ The geometric routing algorithms we
discuss are VOID-1 and VOID-2.  They are rather straightforward
extensions of algorithms FACE-1 and FACE-2 respectively. The latter
two are planar graph geometric routing algorithms presented by Bose et
al \cite{Bose01}.  FACE-1 has a better worst-case performance. However
FACE-2 is more efficient in practice. Hence, we describe VOID-2 in
detail and mention briefly how VOID-1 is designed. Our algorithms are
based on the following observation.

\begin{proposition}
Let $p_1$ and $p_2 \neq p_1$ be two arbitrary points on the edges or
vertices of a geometric graph. Let $V$ be the void such that $p_1$
lies on its border and the line segment $(p_1,p_2)$ intersects $V$. If
there is a point $p_3$ where $(p_1,p_2)$ intersects the border of $V$
then $|p_3,p_2| < |p_1,p_2|$.
\end{proposition}

\begin{figure}
\center \epsfig{figure=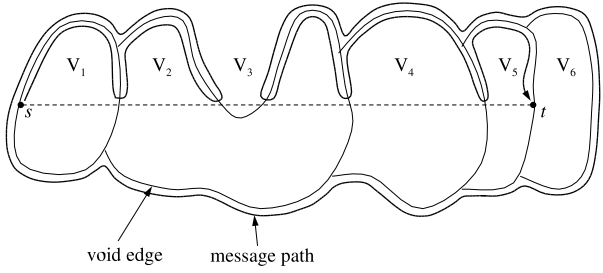,width=8cm,clip=}
\caption{Example route of VOID-2.}
\label{FigVoid2Pic} 
\end{figure}

%
% might need it in the future.
%
% \begin{figure}
% \begin{tabbing}
% 1234\=1234\=1234\=1234\=12345\=12345\=12345\=12345\=12345\=12345\=\kill
% \textbf{Algorithm}: VOID-1 \\
% \>$p_1 \leftarrow s$,\ $p_2 \leftarrow t$,\ $p_3 \leftarrow p_1$ \\
% \>\textbf{repeat} \\
% \>\>/* let $V$ be the void with $p_1$ on  \\
% \>\> \ \  its boundary that intersects $(p_1,p_2)$ */\\
% \>\>\textbf{repeat} \\
% \>\>\>traverse $V$ until finding      \\
% \>\>\>an edge containing point $p_4$  \\
% \>\>\>intersecting $(p_1,p_2)$ \\
% \>\>\>\textbf{if} $|p_4,p_2| < |p_3,p_2|$ \textbf{then} \\
% \>\>\>\>          $p_3 \leftarrow p_4$ \\
% \>\>\textbf{until} $p_4 = p_1$ /* end traversal of $V$  */ \\ 
% \>\>$p1 \leftarrow p_3$ \\
% \>\textbf{until} $p3=p2$ /* target reached */ \\
% \end{tabbing}
% \end{figure}

\begin{figure}[h]
\begin{tabbing}
1234\=1234\=1234\=1234\=12345\=12345\=12345\=12345\=12345\=12345\=\kill
\textbf{Algorithm}: VOID-2 \\
\>$p_1 \leftarrow s$ \\
\>$p_2 \leftarrow t$ \\
\>\textbf{repeat} \\
\>\>/* let $V$ be the void with $p_1$ on  \\
\>\> \ \  its boundary that intersects $(p_1,p_2)$ */\\
\>\>traverse $V$ until reaching      \\
\>\>an edge containing point $p_3$  \\
\>\>intersecting $(p_1,p_2)$ \\
\>\>$p_1 \leftarrow p_3$ \\
\>\textbf{until} $p_1=p_2$ /* target reached */ \\
\end{tabbing}
\caption{Pseudocode for VOID-2. VOID-2 uses void 
         traversal to guarantee delivery.}
\label{FigVoid2Alg}
\end{figure}

Both VOID-1 and VOID-2 sequentially traverse the voids that intersect
the line $(s,t)$. Refer to Figure \ref{FigVoid2Pic} for illustration.
While traversing a void, VOID-1 and VOID-2 look for an intersection
point with line $(s,t)$. Observe that there may be multiple such
points. For example, void $V_2$ in Figure \ref{FigVoid2Pic} has four
such points. VOID-1 and VOID-2 differ in their actions when an
intersection point is encountered. VOID-1 traverses the whole void,
selects the intersection point that is closest to $t$, moves the
message to this point and switches to the traversal of the next
void. For example, VOID-1 would traverse $V_2$ completely and switch
to $V_4$. VOID-2, on the other hand, switches to the next void as soon
as the first intersection point is found. The pseudocode for VOID-2 is
given in Figure \ref{FigVoid2Alg} and an example route that VOID-2
selects is shown in Figure \ref{FigVoid2Pic}.

On the basis of Theorem \ref{TraversalOK} and the discussion above
we state the following.

\begin{theorem}\label{VoidsOK}
Given an arbitrary geometric graph, an intersection semi-closed
neighborhood relation and an arbitrary pair of nodes in this graph,
both VOID-1 and VOID-2 eventually deliver a message from the first
node to the second.
\end{theorem}

\ \\ \textbf{GVG.} As in GFG, void traversal is only needed for a
message to leave a local minimum. Otherwise greedy routing can be
used. Care must be taken to avoid a livelock when message returns to
the same local minimum.  The complete algorithm --
\emph{greedy-void-greedy} (GVG) works as follows. The message is at
first forwarded according to greedy routing. When a node discovers
that it is a local minimum, it notes the distance to target and the
routing switches to VOID-1 or VOID-2.  The routing switches back to
greedy if the distance to target is less than that of the local
minimum.  The process repeats if necessary.  Observe that, according
to Theorem \ref{VoidsOK}, the delivery of a message by either VOID-1
or VOID-2 is guaranteed. Hence, either GVG eventually switches to
greedy routing or delivers the message to the target. Hence the
following theorem.

\begin{theorem}
Given an arbitrary geometric graph, an intersection semi-closed
neighborhood relation and an arbitrary pair of nodes in this graph,
GVG eventually delivers a message from the first node to the second
one in that pair.
\end{theorem}

% \textbf{Application to Unit-Disk Graphs}
%
%
%
% \begin{lemma}
% A unit-disk neighborhood is incident edge closed if every node $a$
% knows the coordinates of each node within $2u$ of $a$; the
% neighborhood is semi-closed if $a$ knows of each node within distance
% $2u/\sqrt{3}$ of $a$.
% \end{lemma}

% \noindent
% \textbf{Implementation and optimizations}: Then the process needs to
% know the neighbors (within $u$) and some neighbors within ``that''
% little sliver. Btw, the longest distance in the sliver is
% $u\sqrt{2}$. Actuality, since the nodes in the sliver can also be
% distributed between the two nodes adjacent to the edge the distance is
% $2u/\sqrt{3}$.
% 
% If each node knows the location of the nodes within $2u$ from it then
% the following is true.
% 
% \indent
%    claim: for each incident edge, the node knows every edge that intersect
%    the incident edge. 

% Observe that one of the nodes of the intersecting edge is within $u$
% of the node.  Hence the message can be sent directly to it. Thus the
% algorithm can be simplified to eliminate the transmission of
% the message over the border edge.

\section{Performance Evaluation}
\label{SecEval}

Note that the greedy phase of GFG and GVG is the same. Hence we only
compare the performance of local minimum avoidance parts: void and
face traversal. We use FACE-2 and VOID-2 for comparison as they are
more efficient in practice then FACE-1 and VOID-1.

\begin{figure} 
\center 
\epsfig{figure=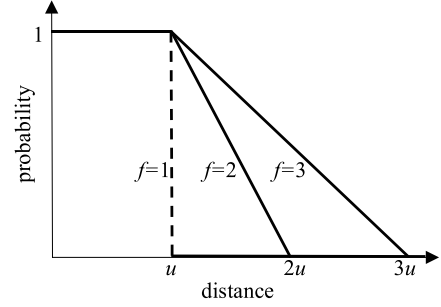,width=6cm,clip=}
\caption{Edge existence probability with respect to the distance
         between nodes.}
\label{figProbability}
\end{figure}

As a primary metric we compare the length, in the number of hops, of
the routes selected by face and void traversal algorithms. Note that a
void traversal algorithm makes routing decisions on the basis of
larger neighborhood information. Therefore, we compared the individual
node memory requirements for face and void traversal algorithms as
well.

\ \\ \textbf{Graph Generation.}  We used randomly generated sets of
graphs with 50, 100 and 200 nodes in a fixed area of 2 by 2 units. The
nodes were uniformly distributed over the area. We used a rather
simple model for fading effect of radio reception.  For each set of
graphs we selected the connectivity unit $u$ to be $0.3$, $0.25$ and
$0.2$ respectively. We also selected a fading factor $f$ to be either
$1$, $2$ and $3$.  The connectivity between nodes was determined as
follows. We deterministically added an edge for every pair of nodes
that were no more than $u$ away from each other.  See Figure
\ref{figProbability}.  If the distance between two nodes was between
$u$ and $f\cdot u$, the edge between them is added
probabilistically. The probability linearly decreased from $1$ to
$0$. Notice that when the fading factor $f$ is equal to one, the
random graphs that we generated were unit-disk.

\begin{figure*}[ht]
\begin{minipage}[b]{.47\linewidth}
\epsfig{figure=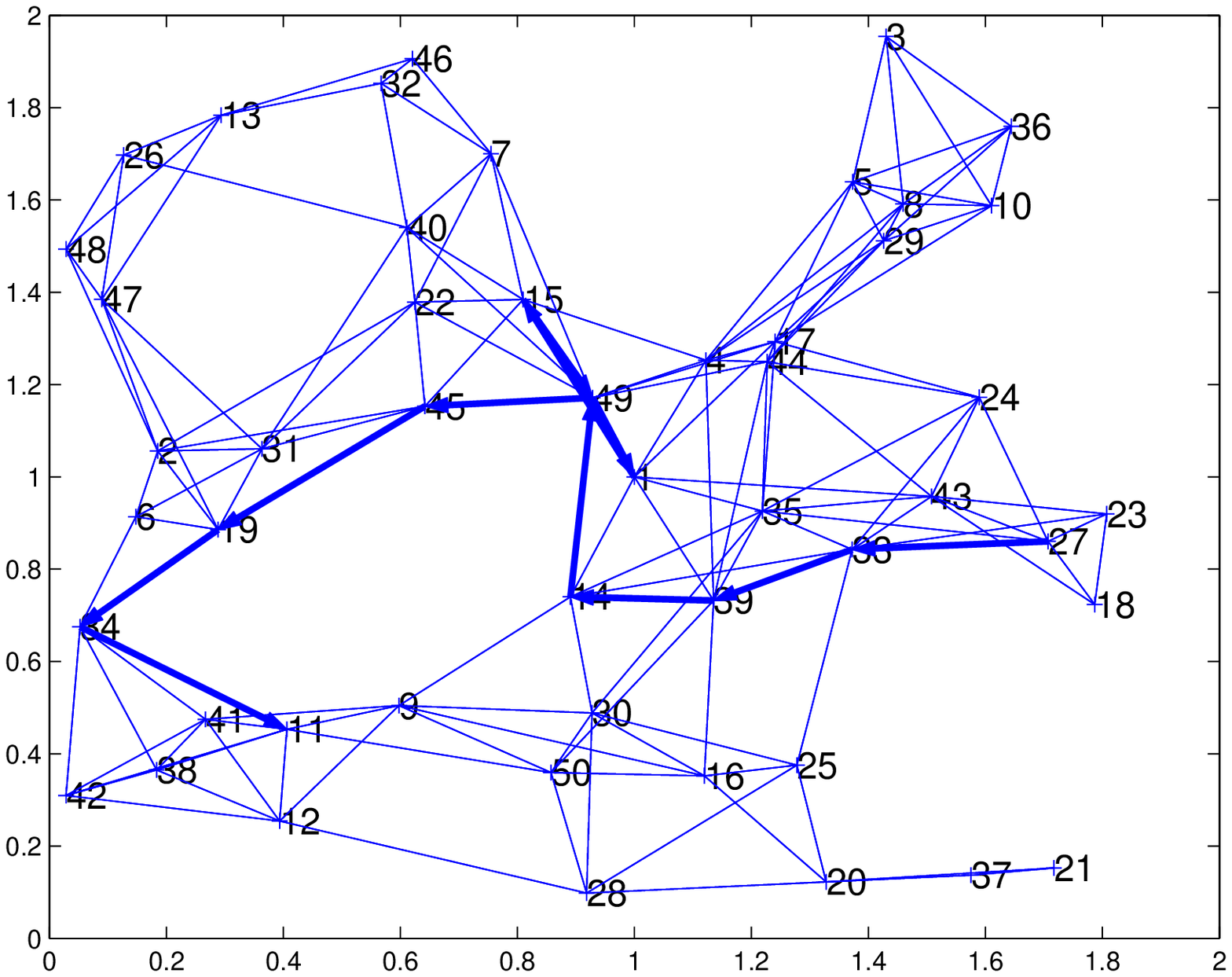,width=\linewidth}
\end{minipage}\hfill
\begin{minipage}[b]{.47\linewidth}
\epsfig{figure=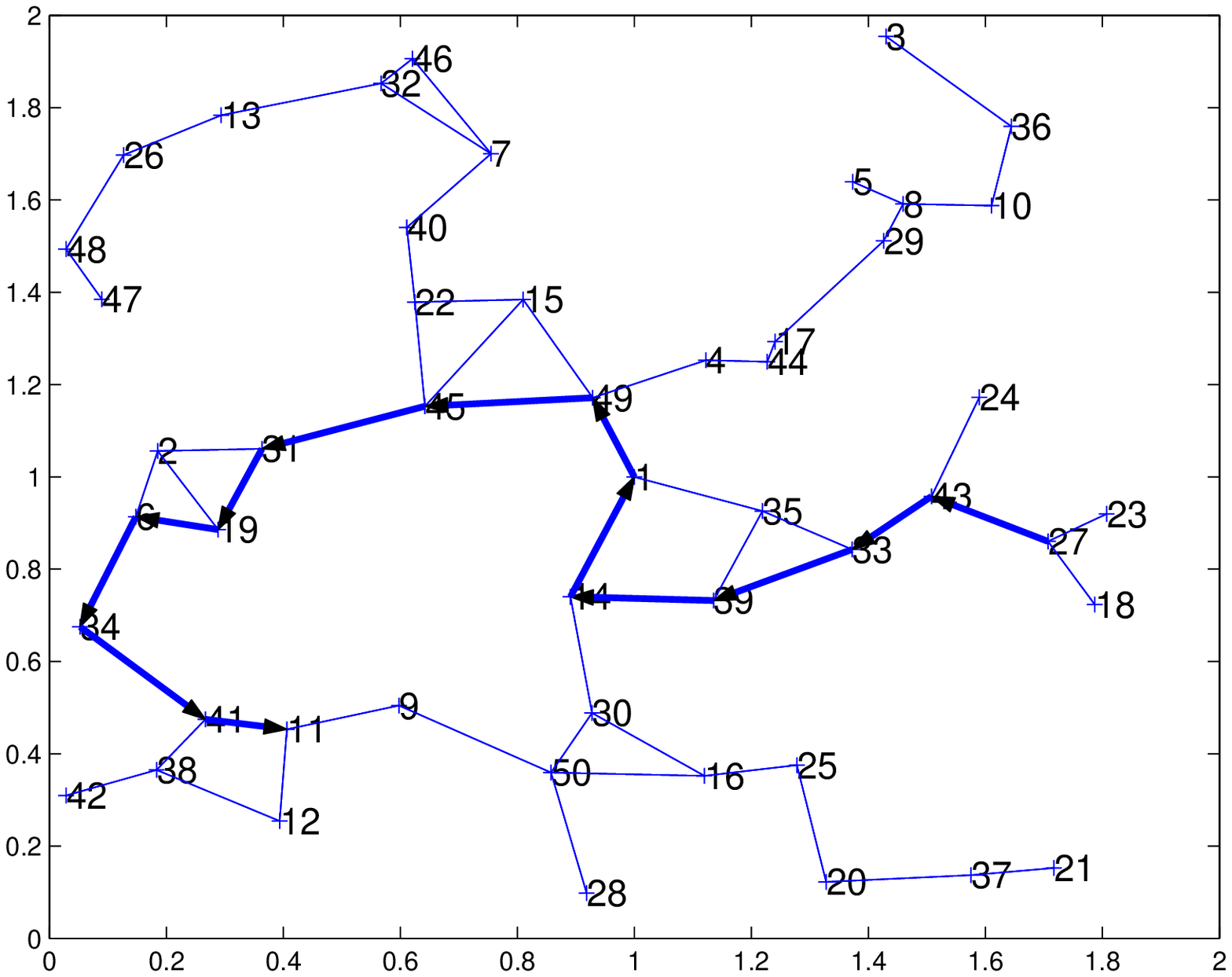,width=\linewidth}
\end{minipage}
\caption{Example routes selected by VOID-2 and FACE-2 between nodes 27
and 11 in a 50-node graph and its unit-disk based subgraph respectively.
The fading factor $f$ is 2.}
\label{figRoutes}
\end{figure*}

In order to compare FACE-2 and VOID-2 we needed to compute unit-disk
subgraphs of the the generated graphs. However, frequently the
subgraphs were disconnected: out of 350 of randomly generated 50-node
graphs with $f$ either $2$ or $3$, only a single graph had a connected
unit-disk subgraph. These subgraphs had to be discarded as unsuitable
for FACE-2. To make the performance comparison, we adopted a different
graph selection strategy. We generated random, connected unit-disk
graphs and added extra links according to the connectivity rules
above. Notice that this strategy favors FACE-2 as it discounts the
graphs that are not usable by FACE-2.

We generated 5 graphs for each number of nodes and fading factor.
This yielded the total of 45 graphs and their planar subgraphs.

\ \\ \textbf{Route Length Comparison.} We implemented the FACE-2 and
VOID-2 algorithms in Java and Matlab. We generated 10 random pairs of
nodes for each set of graphs and used VOID-2 and FACE-2 to compute the
routes between the nodes in each pair. VOID-2 used the original graph
and FACE-2 --- its unit-disk based planar subgraph. The
example routes are shown in Figure \ref{figRoutes}.

\begin{figure}[t]
\center \epsfig{figure=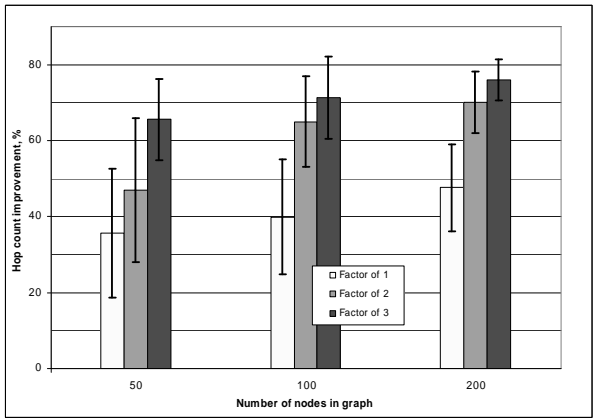,width=8cm,clip=}
\caption{Route length improvement from FACE-2 to VOID-2.}
\label{figCompare}
\end{figure}

We used paired comparison to estimate the improvement from FACE-2 to
VOID-2 in the length of the selected routes as follows. For each pair
of source and destination nodes we calculated the difference in the
hop-count between routes and normalized it to the hop count of the
route selected by FACE-2. That is the comparison is based on the
formula: $(HopCount_{FACE} - HopCount_{VOID})/HopCount_{FACE}$. Figure
\ref{figCompare} summarizes the results. Observe that even on
unit-disk graphs ($f=1$), VOID-2 generates routes that are 35-45\%
shorter than the routes selected by FACE-2. The relative performance
of VOID-2 further improves as the fading factor and density of the
graphs increases.

\ \\ \textbf{Memory Usage Comparison.} In both VOID-2 and FACE-2, each
node $v$ stores the information about its neighborhood nodes $N(v)$.
The size of the neighborhood differs in the two algorithms. In
evaluating the memory usage, we estimated the average size of the
neighborhood. For the FACE-2, the average size of the neighborhood is
the average degree $d$ of the planar subgraph. In our experiments, the
average size of neighborhood in VOID-2, was found to be $f \cdot d$
where $f$ is the fading factor.

\section{Conclusion}
\label{SecConclusion}

The elegance and efficiency of geometric routing algorithms is
remarkable. Recent advent of large-scale wireless sensor networks
increased the relevance of such algorithms. However, the algorithms'
demand for unit-disk based graph planarity hampered the attractiveness
of geometric routing.  With this paper, we lift this restriction and
help the adoption of geometric routing algorithms as the routing
solution of choice in sensor networks.

\section*{Acknowledgments} We would like to thank Ivan Stojmenovic
of the University of Ottawa and M. Kazim Khan of Kent State University
for their helpful comments.

\suppressfloats[t]

\bibliographystyle{IEEEtran}
\bibliography{geometric,routing}

\end{document}